\documentclass[aps,prl,twocolumn,showpacs,nofootinbib,superscriptaddress]{revtex4}

\begin{document}
\title{On the stability and magnetic properties of surface nanobubbles in water}

\author{\textbf{Siddhartha Sen}\footnote{siddhartha.sen@tcd.ie} }

\affiliation{{CRANN}, \\ \emph{Trinity College, Dublin 2, Ireland}}

\author{\textbf{Kumar S. Gupta}\footnote{kumars.gupta@saha.ac.in} }

\affiliation{{Saha Institute of Nuclear Physics, Theory Division}\\ \emph{1/AF 
Bidhannagar, Kolkata 700 064, India}}

\begin{abstract}
A model for gas nanobubbles  is proposed in which their
 remarkable stability is explained as due to the presence of a 
 qualitatively different form of water covering the nanobubble
 surface  which leads to a reduction of the diffusion
 coefficient by a factor of $10^{9}$.  It is shown that this new form of water 
is created by the interaction of the electrons
of water molecules with the zero point vacuum electromagnetic field.
 The model gives
an estimate for the life time of surface nanobubbles, explains
why they are not influenced by surfactants and predicts that
they should exhibit nonlinear paramagnetism.
\end{abstract}


\maketitle



Nanobubbles with radii in the range 25 to 1000 nanometres and contact angles
in the range $135^{\circ}$ to $ 175^{\circ}$ have been observed 
to form on hydrophobic surfaces, where they are remarkably stable 
\cite{ducker,lohse1,seddon,jin,lohse2}.  For bubbles of this 
size lifetimes of the order microseconds are expected due to the 
high Laplace pressure, $\frac{2\gamma}{\rho}$, where $\gamma$ is the surface tension of the liquid and $\rho$, the radius of curvature of the bubble, inside them. This high pressure should drive the gas into the liquid by diffusion and make the nanobubbles unstable.  However the nanobubbles observed are very stable with lifetimes of hours or even days. These bubbles
have unusual properties. For example it is observed that the 
addition of surfactants does not influence their long life 
time and stability. Normally one would expect surfactants,
 which lower the surface tension, to  decrease the bubble lifetimes.

Any theoretical model for nanobubbles should explain
 these basic observational features. A number of ideas 
 have been proposed \cite{ducker,lohse1,seddon,jin,lohse2}. 
Here we would like to propose a very different model \cite{sentalk} 
which can qualitatively
account for the surface  nanobubble properties listed.
 Our model gives an estimate of their  lifetimes,
 explains why they expel surfactants 
and it also  predicts that the surface nanobubbles
 should have nonlinear response to external magnetic fields with
 a saturation value for its magnetic moment per water molecule, of 
 $g\mu_{B}$, where $\mu_{B}$ is the Bohr magneton and $g \approx 10^{-3}$.
 The magnetic moment of the nanoshell is estimated to be 
$\approx 10^{6} \mu_{B}$.

The basic premise of the model is that interfacial water has two phases. The
first phase is normal water. The second phase  is water in the form 
of coherent nanoscale domains of volume $V_c$ containing a suitable number $N$ of electrons associated with the water molecules.
We suggest that it is this coherent phase that is responsible for the
 stability and existence of surface nanobubbles. 

We show that the zero-point fluctuating vacuum electromagnetic (EM)
 field, required to exist by quantum theory,
 can spontaneously generate nanoscale structures.
It can  produce a time-independent 
shift in the position of  orbital electrons \cite{itz}. For the case
 of the hydrogen atom,
this positional shift has been used to provide a good estimate of
 the observed Lamb 
shift of its  $^2S$ energy level \cite{itz,lamb,welt}.
The zero-point  vacuum EM field  also gives rise to the Casimir effect \cite{cas}. 
For two parallel plates close to each other the energy of the vacuum EM field 
between the plates changes if the distance between the plates
is changed. This gives rise to the Casimir force, which has been measured \cite{most}.

In this work, we combine the idea of a universal shift in 
the position of an electron due to the fluctuating vacuum EM 
field with the idea that a change in the volume of a region containing zero point
EM fields gives rise to forces. We apply these ideas to a well defined  nanoscale 
volume  containing a number of electrons and show how an induced EM
force is generated. This interaction lowers the energy of the ground state of 
 the cluster and thus  leads to the formation of a stable spontaneously created 
 coherent  many-electron nanoscale structure. Here we suggest that such
a coherent structure is formed on the surface  of nanobubbles in water
 and is responsible for their remarkable stability.

Let us sketch how this comes about. Our system is the surface of a nanobubble, 
which is assumed to have a well defined surface layer volume $V_c$.
 This volume contains water molecules with $N$ orbiting electrons on
 which the zero point EM field acts.
Furthermore  each water molecule is assumed  to be in one of 
 two electronic states, 
a ground state of energy  $\hbar \omega_1$ and an excited state of
 energy  $\hbar \omega_2$, which is assumed to be stable. 
This assumption is made only for the sake of simplicity. 
It is not an essential requirement of the model. Recent theoretical work on water suggests
that the excited states of bulk water are unstable leading to
 the dissociation of electrons from the water
 molecules on a timescale of femtoseconds \cite{shan}. 
Thus nanoscale structures in the bulk, even if formed are unstable. However this 
might not be the case at the surface or for water in the presence of surfaces 
or other substrates, such as biomolecules.  Here surface effects
can stabilize excited states. We assume this to be the case. 
For instance  biomolecules provide a convenient scaffold for 
nanostructures and there is indeed evidence of different form of 
water adjacent to biomolecules \cite{pollack,senbio}. 
Our  analysis will show that coherent nanoscale structures 
do form as a volume layer at the surface of nanobubbles.  The electrons of
the water molecule are coherent as they all oscillate with a collective
common frequency $\Omega$, different from the 
transition frequency $\omega=\omega_2-\omega_1$ of the water
molecule.  This collective behaviour  arises because, as we show,
the coherent structure has lower energy.  Furthermore a result of
 Frohlich \cite{froh} implies that two oscillating dipoles interacting through
 a compatible oscillating  EM field attract one another while
 molecules or atoms which do not satisfy such a resonance
 condition repel \cite{froh2}.  This explains why surfactants do not
 influence stability: the surfactant molecules, not in
 resonance with the coherent  surface volume molecules are expelled
 by the Frohlich force. Thus gases in the coherent water layer, not 
 in resonance,  are  expelled from the water to the hydrophobic surface,
 leading to the formation of  surface nanobubbles. 

The formation of a coherent water layer
 also prevents diffusion of gases through them. This is 
 because the scattering cross section of gas molecules with a coherent structure
 is greatly increased leading to a decrease in the diffusion coefficient $D$.
 Recall that  $D \approx\frac{<v>}{n\sigma}$, where $<v>$ is
 the thermal speed of the diffusing molecule,
 $n$ is the number density of  water molecules from which it scatters and $\sigma$
 is the scattering cross section of water for the gas molecule concerned. 
 If the surface water contains $N$ water molecules scattering incoherently,
 the scattering is proportional to $N$ while if the scattering is coherent,
 it is proportional to $N^2$ \cite{coey}. The scattering is expected
 to be coherent for the coherent nanoscale domains so  that the effective cross section for scattering is  increased by a factor of $N$ and the diffusion constant is decreased
 by the same factor. We will show that for a coherent domain to 
form $N\approx 10^{9}$. This leads to an increase in the lifetime for gas diffusion
 by a factor of $10^9$ which explains why surface nanobubble have
 long lifetimes.  The lifetime is expected to be further enhanced by the Frohlich repulsive force  between non resonating molecules. 
 
As emphasized already, an essential assumption of our work in that a nanoscale
 collection of  electrons exist within a physical
 volume $V_c$ which can be clearly identified. For the present
 work, the volume $V_c$ is identified with the
 surface volume of nanobubble in water which contains a collection
 of $N$ electrons. The vacuum electromagnetic field couples
 to all these electrons, each of which has mass $m$ and charge $e$. 
Let $\vec{E}(t,\vec{x})$ denote the time dependent
 electromagnetic field due to the vacuum fluctuation
 and $\vec{\delta_i}$ denote the fluctuation in the position of
 the $i^{\mathrm {th}}$ electron due to the effect of $\vec{E(t,x)}$. Then we have
\begin{eqnarray}
m \sum_i\ddot{\vec{\delta_i}}(t, \vec{x}) &=& N e \vec{E} (t, \vec{x}),~~~~i = 1,2,.....N.
\end{eqnarray}
Taking the Fourier transform, we get
\begin{equation}
\label{ft}
-m \sum_i\omega^2 \vec{\delta_i}(\omega, \vec{x}) = N e \vec{E} (\omega, \vec{x}),
\end{equation}
where we have used the same symbols $\vec{\delta}$ and $\vec{E}$ 
for the Fourier transforms of these quantities. Thus we get
\begin{equation}
\left | \sum_i{\vec{\delta}_i} (\omega, \vec{x}) \right |^2 = N^2 \frac{e^2 E^2 (\omega, \vec{x})}{m^2 w^4},
\end{equation}
where $E^2 \equiv |{\vec{E}}|^2$.

We now calculate the time average of the fluctuations. 
We assume that the fluctuations are independent so that the 
cross terms vanish. We also assume that the time
 averaged fluctuation $<|\vec{\delta_i}|^2> \equiv <\delta^2>$ are 
the same for all the electrons. With these assumptions and using 
 the result for the value of $\delta^2$ given in \cite{itz} we get our final expression for the average squared fluctuation of the position of an electron when it is part of an assembly of $N$ electrons as
\begin{equation}
\label{delta}
< \delta^2 >_{N} = 2 N  \alpha \lambda_c^2 \ln{\frac{1}{\alpha^2}} \equiv \delta_N^2,
\end{equation}
where $\alpha = \frac{e^2}{\hbar c}$ is the fine structure 
constant and $\lambda_c = \frac{\hbar}{m c}$ is the
 Compton wavelength of the electron.
 The factors in the log term come from
limits of $\omega$ integration which are taken 
to be $\frac{mc^2}{\hbar}$ and $\frac{me^4}{\hbar^3}$ 
representing a relativistic cut off set by the electron
 mass and a lower frequency cut off set by the atomic scale 
 Bohr frequency \cite{itz}. We will set
$\ln\frac{1}{\alpha^2} \approx 8$. Thus $<\delta^2>_{N} \approx 16N \alpha \lambda^2_{c}$. This is a constant universal expression. 

We now use this result to determine the coupling of a charge to the zero point
field when it is part of an assembly of charges close together.
The idea we use is that the fluctuation-induced position change $\delta_{N}$ produces a 
change of the well defined nanovolume $V_c \approx l^3$ and this
leads to a force. The volume of the nanoshell is
set by the wavelength $l$ corresponding to a transition 
of energy $\approx 10$ eV.
This induced force tells how the vacuum EM field 
interacts with a charge belonging to an assembly.

To determine this force.
we start with our well-defined nanocluster of charged particles
in volume $V_c$. The energy density of the fluctuating field in this volume
is $|\vec{E}|^2= \frac{1}{2V_c}\hbar\omega$ for frequency $\omega$, where $\vec{E}$ is the vacuum EM field, given by 
$\vec{E} = \vec{u}\frac{\sqrt{\hbar\omega}}{\sqrt{2V_c}} e^{-i\omega t}$. Consider the effect of volume change due to the position change 
$\sqrt{<\delta^2(t,\vec{x})>_{N}} \approx 4\sqrt{\alpha \lambda_c^2 N}$ on $\vec{E}$
The volume change produces, as we now show, an induced EM force,
 which we write as $e\vec{E}_{f}(t)$. 

 We determine $\vec{E}_{f}(t)$ in two steps. In the first step we define $\vec{E}_{f}$
by the equation 
\begin{eqnarray}
\sqrt{\alpha} \vec{E}_{f}(t)&=&\vec{u}\sqrt{\hbar\omega}[\frac{1}{\sqrt{2(V_c+\delta V)}}-\frac{1}{\sqrt{2V_c}}]e^{-i\omega t}
\end{eqnarray}
where $\frac{\delta V_c}{V_c}=\frac{3\sqrt{<\delta>^2}}{l}$. A constraint to remember is that we must have $\frac{\delta V}{V}<<1$.
Thus  we get
\begin{equation}
e\vec{E}_{f}(t)=\vec{u}~\frac{3}{2}\sqrt{16c\frac{N\alpha \lambda_c^2}{2V_c\omega}}(\frac{1}{l})\hbar\omega~e^{-i\omega t}.
\end{equation}
The fluctuating EM force  $e\vec{E}_{f}$ can now be used to  produce
 an  interaction energy term,  $e\vec{E}_{f}(t).\vec{x}$ with an electron,
 located at  $\vec{x}$ which is simply the usual $\frac{e}{c}~\vec{j}.\vec{A}$ term, where the vector potential $\vec{A}$ is defined by 
$\vec{E}_{f}=\frac{1}{c}~\frac{\partial \vec{A}}{\partial t}$. 
 Thus the induced EM term constructed is a standard field-current interaction and
 can cause a transition between states $|i>, |f>$. Its transition matrix element
 is given by the expression
\begin{eqnarray}
<i|e\vec{E}_{f}.\vec{x}|f>&=&  \frac{<i|(\vec{x}.\vec{u})|f>}{\sqrt{2 V_c  \omega/c}}\frac{3}{2}\sqrt{\alpha} \sqrt{\frac{16N\lambda_c^2}{l^2}}~\hbar\omega
\end{eqnarray}
where $ \vec{x}.\vec{u}=rl \cos\theta$ , and  $\cos\theta$ is the angle
 between the vectors $\vec{x}, \vec{u}$.  The expression for the 
 vacuum field induced transition amplitude represents a coupling 
between two electronic states and the zero point photon induced em field.
It has the structure of the usual dipole transition but with a multiplicative
factor $\sqrt{\frac{16N\lambda_{c}^2}{l^2}}$ due to 
the fact that the interaction is induced from zero point fluctuations.
When $N$ is $\approx 10^{9}$ this factor is of order unity and the mixing
of states takes place with high probability. This is what we think happens
in the nanobubble shell.

Another important result that can be extracted from the transition
matrix element  is a frequency relation  between a collective frequency $\Omega$
and the transition frequency $\omega$, which is,
\begin{eqnarray}
\Omega &=&G\omega,\\
G&=&\sqrt{\alpha}~r~\frac{3}{2}~\sqrt{\frac{Nc}{2V_c\omega}}~\left ( \sqrt{\frac{16\lambda_c^2}{l^2}} \right ).
\end{eqnarray}
We note that $\frac{<i|e\vec{E}_{F}.\vec{x}|f>}{<i|\vec{x}.\vec{u}|f>}$
 is a characteristic energy associated with the transition which we have
written as $\hbar \Omega$. We will show, shortly, that this energy
term, representing the mixing
 between the ground state and the exited state, lowers the ground state energy.
 Hence we expect that all the atoms in our nano assembly will have 
this frequency $\Omega$. The value of $G$ is $\approx 10^{-3}$ for 
$N \approx 10^{9},r\approx 3\times 10^{-8}$ cm, $\omega \approx 10^{16} s^{-1}$, and $V \approx 10^{-15}$ cc.
 
   A result of this form was first
 derived  Preparata et al in \cite{prep1} for water molecules using a path integral
  quantum field theory  approach. They 
 pioneered  the idea that nanoscale structures would  form driven by 
time dependent  zero point em fields \cite{prep3}.
 However our approach  is different from that of Preparata 
 et al and one of our basic  results is a crucially different 
from those derived  in \cite{prep3}, namely our linking equation
 between the collective oscillation $\Omega_{ij}$ and the 
transition  oscillation  frequencies $\omega_{ij}$ has an 
additional numerical factor of $\sqrt{\frac{16N\lambda_{c}^2}{l^2}}$ present. 
This factor appears for the physical reason that we have explained. 

We now show that the mixing of states can lower the ground state 
energy by just considering frequencies. Note the vacuum 
induced electric field is  independent of $\vec{x}$ and small.
 Hence a simple perturbation treatment of its effect is allowed.
 We describe energies in terms
 of frequencies $\Omega, \omega_1, \omega_2$, neglecting numerical factors. 
 Consider the Hamiltonian 
\begin{eqnarray}
\label{int}
\frac{H}{\hbar}& =& \left ( 
\begin{array}{cc}
\omega_1 & \Omega  \\
\Omega & \omega_2  
\end{array} 
\right )
\end{eqnarray}
which acts on the states characterized by eigenvalues $\omega_1$ and $\omega_2$ and the interaction term mixes the two states. The eigenvalues of the interaction Hamiltonian $H$ are given by
\begin{equation}
\lambda_{\pm} = \frac{(\omega_1 + \omega_2)}{2} \left [ 1 \pm \sqrt{1 + \frac{4 ( \Omega^2 - \omega_1 \omega_2 )}{(\omega_1 + \omega_2)^2}} \right ].
\end{equation}
For $\Omega^2 >\omega_1 \omega_2$ we see that one of the eigenvalues is
\begin{equation}
\lambda_- \approx - \frac{\Omega^2 - \omega_1 \omega_2}{(\omega_1 + \omega_2)} < 0.
\end{equation}
The condition on $\Omega$ tells us that $\lambda_{-}$ is most negative when
$\omega_2$ is small. This implies  that the energy is  lowered
if the transition is between levels one of which is close to the ionization threshold.
Thus the mixed state is expected to have a loosely bound electron.
In our application the ground state energy is $\omega_1$
 and we assume that the excited state is close to the ionization threshold  
i.e. $\omega_2 \approx 0$. Then, it is clear that the ground state energy
 is lowered due to the mixing and this lowering of the ground state energy 
 is the physical reason  for the formation of coherent nanoclusters.
We stress that although  the  fluctuating EM field mixes the ground state
 with an excited state both states involved 
 are bound states. No photon leaves the system.

Let us work out the magnetic properties expected for the surface nanobubbles
assuming they have a coherent water layer of
 volume $100^3$ cubic nm. This size is fixed by the wavelength of the 
 $\approx 10$eV near ionisation excitation level of water. The current in  the model $j_{e}=e\frac{\Omega}{2\pi}$,
 is interpreted as due to the collective orbiting motion of
  outer electrons of water molecules.  This gives an average 
magnetic moment $\mu$ per water molecule in the coherent domain given by
\begin{equation}
\mu=\frac{\pi r^2 e\Omega}{2\pi c},
\end{equation}
where $r$ is the size of the water molecule and $\Omega$ is 
 the collective frequency of the cluster.
Putting in numbers $r \approx 3 \times 10^{-8}$ cm,
and $\Omega \approx 10^{13}$ we find that $\mu \approx 10^{-3} \mu_{B}$, where
$\mu_{B}$ is the Bohr magneton. We suppose that when a static external magnetic field 
 is introduced it interacts with the  electron currents 
present in the coherent domain. Thus for a coherent domain
interacting with an external magnetic field $\mathbf{B}$,
the static interaction 
is ]$V=N\mathbf{\mu}.\mathbf{B}$.

We now apply our general results to surface nanobubble
water layers, continuing to use the simple model in which the coherent
mixing leads to a lower energy ground state. We  described this ground state in 
terms of two time-dependent coherent oscillatory basis states.
These we now take to be labelled by angular
momentum spherical harmonic labels  $l=0,m=0$ and $l=1,m=0$ that describe the  electrons.
The ground state wavefunction is written as
\begin{equation}
\zeta(\mathbf{\Omega},t)=\gamma_0(t) Y_{0,0}(\mathbf{\Omega})+\gamma_1(t)  Y_{1,0}(\mathbf{\Omega}),
\end{equation}
where $\mathbf{\Omega}$ is the direction of the orbiting electron
angular velocity.

The magnetic field mixes the basis
vectors of the coherent state which
we describe by a mixing angle $\theta$.
Once this step has been taken we can calculate the magnetic moment
of the system, assuming $V$  is a perturbation and as it is static, it does not
modify the time dependence of $\gamma_i(t), i=0,1$.
 The eigenvalues of the system with the magnetic field are,
\begin{eqnarray}
\lambda_1&=&\frac{\omega -\sqrt{\omega^2+4V^2}}{2},\\
\lambda_2 &=& \frac{\omega+\sqrt{\omega^2+4V^2}}{2}. 
\end{eqnarray}
From $\lambda_1, \lambda_2$ the corresponding eigenfunctions 
are constructed.  
In terms of them the ground state wave function becomes
\begin{eqnarray}
 \zeta &=& A Y_{0,0}+B Y_{1,0},\\
A&=&(\gamma_0(t)\cos\theta -\gamma_1(t)\sin\theta),\\
B&=&(\gamma_1(t)\cos\theta+\gamma_0(t)\sin\theta),
\end{eqnarray}
where $\tan\theta=\frac{1-\sqrt{1+x^2}}{x}, x=\frac{2V}{\omega}$. 
Using this expression we now calculate the time average value of the
magnetic moment $P_{Av}$ in the presence of an external magnetic field.
This is simply the time average of $\epsilon_{\mu}.\epsilon_{\mathbf{B}}$
evaluated on the ground state written in terms of the new basis wavefunctions.
Here $\epsilon_{\mathbf{B}}$ and $\epsilon_{\mu}$ are the unit
vectors in the direction of the field and the orbiting electron angular velocity, respectively
 Thus
\begin{equation}
 P_{Av}=\int d\Omega \zeta^{*}(\mathbf{u},t) \cos\theta \zeta(\mathbf{u},t),
\end{equation}
which gives
\begin{equation}
 P_{Av}=-\kappa \sin 2\theta=-\kappa\frac{x}{\sqrt{1+x^2}},
\end{equation}
where $\kappa=\frac{1}{\sqrt{3}}~\frac{\Omega^2-2\lambda^2_0}{\Omega^2+2\lambda^2_0} \approx 0.58$
and $\lambda_{0}=\frac{1}{2}[\omega-\sqrt{\omega^2+4\Omega^2}]$
This is the expression for the expected nonlinear response 
of an outer orbiting electron of a water molecule
that belongs to the surface coherent layer to an external magnetic field.
The value for the total magnetic moment for the surface nanobubble
 is $\frac{N\pi er^2\Omega}{2\pi c} \approx 10^{6} \mu_{B}$, 
where $\mu_{B}$ is the Bohr magneton.
Since this result is for a stable ground state it 
should be temperature independent. 

In conclusion, we have suggested that surface nanobubbles are
 stable because of a new structured phase
of water, first suggested by Preparata and coworkers using 
the methods of quantum field theory. Our treatment is based on simpler ideas 
of fluctuations and give different  results. 
The main difference is that our model suggests that nanoscale structures
need to have a well defined starting volume. We do not expect them in the bulk
but for surface nanobubbles where the volume at the surface is well defined.
The second difference is that our induced em field has an additional factor
representing the vacuum fluctuating origin of the force.

Surface nanobubbles can only have sizes
that are compatible with a fixed coherent volume of $\approx 100^3$ cubic nm.
They are expected to be charged, due to the mixing of the ground state with
a state close to the ionization threshold, they should expel surfactants due
to quantum forces that repel molecules  that are not in resonance, which
 explains why surface nanobubbles form,  
while the increase in the scattering cross section by a factor of $N \approx 10^{9}$
due to the formation of coherent domains, explains their long life. 
Finally we showed that they are expected to exhibit nonlinear orbital paramagnetism.
The model described has  been used to explain the observed relatively long
lifetime of microbubbles\cite{coey} and the observed temperature independent nonlinear magnetic properties of doped Cerium Oxide \cite{coey2}  

\section*{Acknowledgement}

KSG and SS would like to thank Prof. Alvaro Ferraz (IIP-UFRN-Brazil)
for the hospitality at IIP-Natal-Brazil, where this work was carried
out. SS would like to thank Prof Michael Coey of CRANN for getting him interested
in nanobubbles, for many helpful discussions of experimental results and for his useful comments about the paper, especially regarding the significance of Eq (21).


\begin{thebibliography}{99}
\bibitem{froh2}  The reason for an attractive resonance dipole dipole
 interaction  can be understood by considering, following Frohlich, a model
 of free dipoles interacting with an oscillating electromagnetic field.  The interaction
Hamiltonian H is taken to be 
\begin{displaymath}
 H=g\int \psi^{*}(x,t)\psi(x,t) A(x,t)
\end{displaymath}
where $\psi^{*}(x,t)$ represents the dipole creation operator and $\psi(x,t)$ the dipole
destruction operator and $A(x,t)$ the oscillating electromagnetic field
where  we have approximated all fields by scalar fields. The oscillatory character of $A(x,t)$ means we can write it as
\begin{displaymath}
 A(x,t)=\frac{1}{\sqrt{V}} \sum [a_q e^{-1\omega_q t+\mathbf{q}.\mathbf{x}}+ hc]
\end{displaymath}
This gives rise to a positive interaction of the form
\begin{displaymath}
 V_q \approx \frac{\omega_q}{(\epsilon_p-\epsilon_{p+q})^2-(\omega_q)^2}
\end{displaymath}
where $\epsilon=\epsilon_p-\epsilon_{p+q}$ is  energy difference between the two dipoles and $\omega$ is the photon energy.
Integrating $\omega$ over the allowed momenta values of the photon
with an appropriate physically determined  cut off, $\Omega$,  gives the potential energy as a function of the difference in energies of the two dipoles. In the simple model we
take the dipoles to be free air molecules in the self generated gas cavity that surrounds a hydrophobic site. The quantum fluctuating momentum scale set by the quantum uncertainty principle is for coherent domain.  Carrying out the momentum integration gives the dipole- photon potential $V_{d,\gamma}(\epsilon)$ to be
\begin{displaymath}
V_{d,\gamma}(\epsilon) \approx (\frac{g^2}{\Omega})(\frac{\epsilon}{2\Omega}\ln|\frac{1+\frac{\epsilon}{\Omega}}{1-\frac{\epsilon}{\Omega}}|-1)
\end{displaymath}
where $\epsilon=\epsilon_1-\epsilon_2$ the  difference of energies of the two dipoles and $\Omega$ is the photon energy corresponding to the thermal
 momentum cutoff momentum. We see this potential is attractive for
$\epsilon \leq \Omega$ and repulsive otherwise.  We note that this attractive potential, and its associated attractive force, increases as $\epsilon \Rightarrow 0$ which corresponds to the oscillators being in resonance. This new force, of quantum origin, only appears when the dipoles are oscillating and the electromagnetic field is time dependent. The oscillation of the dipoles in the model come from the fact that they are confined to a coherent domain and 
are represented by oscillating waves.  This new quantum force  brings resonating dipoles close together.
\end{thebibliography}
\end{document}